# Light scattering in disordered honeycomb photonic lattices near the Dirac points

Yaroslav V. Kartashov,[1,2,*] Julia M. Zeuner,[3] Alexander Szameit,[3] Victor A. Vysloukh,[4] and Lluis Torner[1]

[1]*ICFO-Institut de Ciencies Fotoniques and Universitat Politecnica de Catalunya, 08860 Castelldefels (Barcelona), Spain*
[2]*Institute of Spectroscopy, Russian Academy of Sciences, Troitsk, Moscow Region, 142190, Russia*
[3]*Institute of Applied Physics, Friedrich-Schiller-Universität Jena, Max-Wien-Platz 1, 97743 Jena*
[4]*Departamento de Fisica y Matematicas, Universidad de las Americas, Santa Catarina Martir 72820, Puebla, Mexico*

We address Anderson localization in disordered honeycomb photonic lattices and show that the localization process is strongly affected by the spectral position of the input wavepacket within the first Brillouin zone of the lattice. In spite of the fact that in regular lattice the expansion of the beam is much stronger for the excitation near the Dirac points, where light exhibits conical diffraction, than for the excitation at the center of the Brillouin zone, where the beam exhibits normal diffraction, we found that disorder leads to pronounced Anderson localization even around the Dirac points. We found that for the same disorder level the width of the averaged output intensity distribution for excitations around the Dirac points may be substantially larger than that for excitations at the center of the Brillouin zone.

Light transport in disordered photonic lattices has [1] received much attention in recent years, starting with the seminal ideas reported in [2]. Following first experiments [3], the breakthrough came with direct observation of Anderson localization in two- [4] and one-dimensional photonic lattices [5]. Various aspects of Anderson localization were analyzed at lattice interfaces, in lattices with second-order coupling, and for different disorder models [6-12]. However, all of these works deal with systems whose dynamics is governed by a Schrödinger equation. At the same time it was shown that in honeycomb photonic lattices – which may be viewed as the optical version of graphene - the evolution of the system for some parts of the band structure can be described by the relativistic Dirac equation that gives rise to numerous intriguing properties [13]. Such photonic graphene is a powerful platform for studying a variety of phenomena, such as non-Hermiticity [14], ultra-strong pseudo-magnetic fields [15], dynamic localization [16], and even topological protection [17]. Only recently has the impact of disorder on the characteristic feature of the band structure of honeycomb photonic lattices been studied [18].

In this Letter we aim to answer an important question: How do the internal structure of the input wavepacket and the position of the center of its spectrum within the first Brillouin zone affect the process of Anderson localization in the presence of disorder in honeycomb photonic lattices? To answer this question, we studied the evolution of Bloch waves with broad Gaussian envelopes taken from the top of the first allowed band, where normal diffraction takes place, and from one of the Dirac points, where conical diffraction occurs, in disordered honeycomb photonic lattices. We found that in the vicinity of the Dirac points, wave localization is considerably weaker.

We describe the propagation of a laser beam along the $\xi$ axis of a material with an imprinted transverse modulation of refractive index by the Schrödinger equation for the dimensionless light field amplitude $q$:

$$i\frac{\partial q}{\partial \xi} = -\frac{1}{2}\left(\frac{\partial^2 q}{\partial \eta^2} + \frac{\partial^2 q}{\partial \zeta^2}\right) - R(\eta,\zeta)q, \qquad (1)$$

where the propagation distance $\xi$ is scaled to the diffraction length; the transverse coordinates $\eta, \zeta$ are scaled to the characteristic beam width; the function $R(\eta,\zeta) = \sum_k p_k \exp\{-[(\eta-\eta_k)^2 + (\zeta-\zeta_k)^2]^2/a^4\}$ describes the shape of the lattice composed from super-Gaussian waveguides of width $a$, whose centers $(\eta_k, \zeta_k)$ are placed in the nodes of the honeycomb grid with separation $d$ between centers of the nearest neighbors and whose depths are described by the parameter $p_k$ [see Fig. 1(a) for the shape of the lattice].

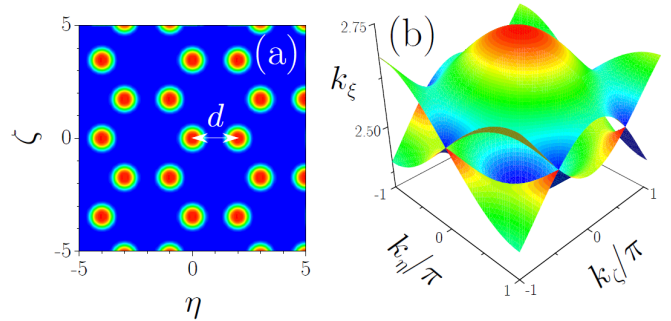

Fig. 1. (a) Honeycomb lattice and (b) first two allowed bands from its Floquet-Bloch spectrum at $p=8$, $a=0.5$, $d=2$.

The Floquet-Bloch spectrum of such a honeycomb lattice is unique due to the presence of so-called Dirac points (see [13] and references therein) where the first and second allowed bands touch each other [Fig. 1(b)]. The unusual structure of the spectrum in the vicinity of Dirac points leads to unusual diffraction properties in the regu-lar lattice. The beam $q|_{\xi=0} = q_C(\eta,\zeta)\exp[-(\eta^2+\zeta^2)/w^2]$ whose shape matches the shape of the Bloch wave $q_C(\eta,\zeta)$ from the top of the first allowed band (the center of the first Brillouin zone with Bloch momentum $k_\eta, k_\zeta = 0$) with a broad Gaussian envelope of width $w$ exhibits normal diffraction upon propagation in the regular lattice, i.e. the envelope of the wavepacket remains bell-shaped at any distance $\xi$ [Fig. 2(a)]. In contrast, the input wavepacket $q|_{\xi=0} = [q_{D1}(\eta,\zeta)+iq_{D2}(\eta,\zeta)]\exp[-ik_\eta\eta - ik_\zeta\zeta - (\eta^2+\zeta^2)/w^2]$ representing a combination of Bloch modes from the bottom of the first $q_{D1}(\eta,\zeta)$ and top of the second $q_{D2}(\eta,\zeta)$ allowed bands in the lattice spectrum and Bloch momentum from one of the Dirac points, for example from the point at $k_\eta = 0$, $k_\zeta = -\pi$ [Fig. 1(b)], exhibits specific "conical" diffraction [13] accompanied by the formation of an expanding bright ring of nearly constant width [Fig. 2(b)]. Notice the difference in the internal field distributions in the inputs leading to normal and conical diffraction. The normally diffracting beam features a honeycomb intensity structure with all waveguides excited in phase, while excitation at the Dirac point features hexagonal, rather than honeycomb intensity distribution and a complex phase structure. These two types of excitation were selected here because in both cases the integral center of the diffraction pattern does not exhibit transverse displacement, which is unavoidable for other points within the Brillouin zone and which can introduce unnecessary asymmetries even in the presence of disorder.

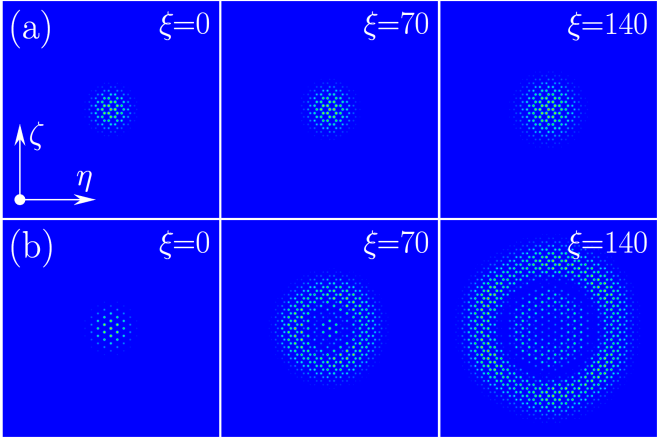

Fig. 2. Intensity distribution at different distances illustrating diffraction in regular lattice for the excitation at the top of the first allowed band (a) and for the excitation in one of the Dirac points (b). The width of the input envelope is $w = 10$.

One can observe from Fig. 2 that despite the equal widths of the input wavepackets the expansion for the excitation in the Dirac point at fixed propagation distance $\xi$ is much more pronounced than the expansion for normal excitation. The natural question arises how different rates of expansion upon normal and conical diffraction affect the process of Anderson localization in the presence of disorder. To answer this question, we introduce diagonal disorder into the honeycomb lattice by letting the depths of individual waveguides $p_k$ fluctuate in each lattice realization in the interval $[p - p_d, p + p_d]$, where $p$ is the ensemble averaged waveguide depth, and the parameter $p_d$ determines the level of disorder in the system.

The distribution of $p_k$ is uniform on the interval $[p - p_d, p + p_d]$. Further, we set $p = 8$ and use waveguide widths $a = 0.5$ and separation $d = 2$. These numbers correspond to the standard parameters of laser-written waveguide arrays that were successfully used before for observation of Anderson localization [8]. We generated $Q = 10^3$ realizations of disordered lattices for each $p_d$ and each width $w$ of the input beam envelope considered, and solved Eq. (1) for each realization up to the distance $\xi = 500$, at which the beams in regular lattice broaden drastically for both normal and Dirac excitations. The averaged output intensity distribution $I_{av}(\eta,\zeta)$ and integral form-factor $\chi_{av}$ were calculated using expressions

$$I_{av}(\eta,\zeta) = \frac{1}{Q}\sum_{i=1,Q}|q_i(\eta,\zeta)|^2,$$
$$\chi_{av}^2 = \frac{1}{QU^2}\sum_{i=1,Q}\int\int_{-\infty}^{\infty}|q_i(\eta,\zeta)|^4\,d\eta d\zeta, \qquad (2)$$
$$U = \int\int_{-\infty}^{\infty}|q_i(\eta,\zeta)|^2\,d\eta d\zeta,$$

where $q_i(\eta,\zeta)$ are the output field distributions for different realizations and $U$ is the input power that is conserved upon propagation. We checked that in all cases, except for very small disorder levels $p_d < 0.03$, the localization regime is achieved at $\xi = 500$, i.e., the initial expansion of the wavepacket nearly stops and the averaged quantities closely approach their asymptotic values determined by the degree of disorder. Notice that the form-factor $\chi_{av}$ is inversely proportional to the width of the most intense Anderson-localized fraction of the output wavepacket (due to fourth power of the field modulus $|q_i(\eta,\zeta)|$ in the integrand in definition of $\chi_{av}$ the small-amplitude background is discriminated and main contribution to the integral comes from high-intensity regions [8]), i.e., larger $\chi_{av}$ implies better localization.

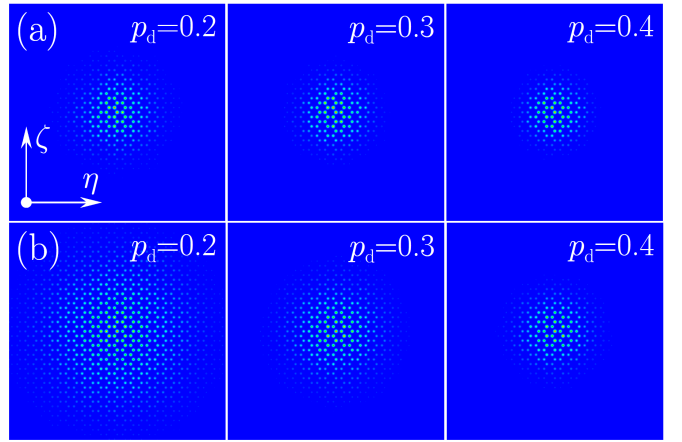

Fig. 3. Averaged intensity distributions at $\xi = 500$ for different disorder levels for the excitation at the top of the first allowed band (a) and for the excitation at one of the Dirac points (b). The width of the input envelope is $w = 10$.

The impact of the level of disorder on the averaged output intensity distributions is illustrated in Fig. 3. With increase of disorder, the initial ballistic spreading for each lattice realization is replaced by the localization. The input beam excites multiple Anderson modes, while the effi-ciency of excitation of each particular mode is dictated by the overlap integral between the mode shape and the input beam. Since different Anderson modes feature different propagation constants, the subsequent beam evolution is dictated by beating between excited modes. The conical diffraction ring is not observable already at $p_d \sim 0.1$. One of the most important results of this Letter is that for small and moderate disorder levels much larger expansion rates for excitations in the Dirac point result in much broader averaged intensity distributions (hence, weaker Anderson localization) in the disordered lattice. The difference in widths of the averaged output intensity distributions for normal and Dirac excitations is most pronounced for the disorder levels $p_d \sim 0.1-0.3$. Generally, growing $p_d$ results in the contraction of the averaged intensity distributions. The difference between normal and Dirac excitations washes out with the increase of $p_d$, since disorder destroys the specific band-gap structure of the honeycomb lattice, and already at $p_d \sim 0.7$ the output intensity distributions for these two types of excitation look very similar. In order to confirm that the localization regime was achieved, we plot in Fig. 4 the cross-sections of averaged intensity distributions $I_{av}$ at $\zeta=0$ in logarithmic scale. Linearly decaying tails visible in these plots confirm exponential localization for both normal and Dirac excitations. The decay rate of the exponential tails rapidly increases with $p_d$.

Remarkably, the comparison of intensity cross-sections in Fig. 4 shows that, while for normal excitation the tails decay slightly faster than for the Dirac excitation, the difference in decay rates even at $p_d=0.3$ is far less pronounced than the difference in widths of the corresponding averaged intensity distributions. This implies that the characteristic decay length for tails of Anderson-localized patterns in the transverse plane is determined mostly by the level of disorder acting in the system, while the output width strongly depends on the position of the input excitation within the Brillouin zone.

As mentioned above, the origin of weaker localization in the vicinity of Dirac point may be connected with the fact that corresponding excitation is spatially chirped, in contrast to the excitation in the center of the Brillouin zone featuring flat phase front. Therefore the excitation in the vicinity of Dirac point experiences faster ballistic spreading at the initial stage of propagation. Even in the presence of disorder, due to faster expansion at the ballistic stage the excitation in Dirac point reaches further waveguides before the onset of localization, that results in weaker overall localization.

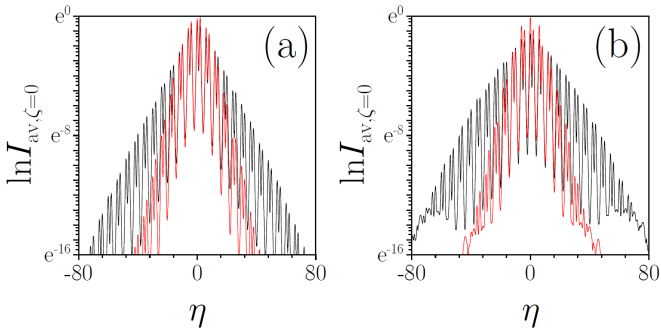

Fig. 4. Cross-sections of averaged intensity distributions at $\zeta=0$ for $p_d = 0.3$ (black curves) and $p_d = 0.8$ (red curves) in the logarithmic scale. Panel (a) corresponds to the excitation at the top of the first allowed band, panel (b) corresponds to the excitation at the Dirac point. The width of the input envelope is $w=10$.

The dependence of the averaged form-factor on the level of disorder at fixed width of the input envelope is shown in Fig. 5(a) for different types of excitation. One can see that the form-factor approaches the same asymptotic value for normal and Dirac excitations for large disorder levels, when the band-gap structure of the lattice spectrum is destroyed. The dependencies $\chi_{av}(p_d)$ are almost linear at $p_d \to 0$, but they have different slopes, giving rise to drastic differences in the widths of the output intensity distributions. Since we are using relatively broad excitations that experience considerable reshaping due to disorder, the dependence $\chi_{av}(p_d)$ may be nonmonotonic. It should be mentioned that the difference in the widths of the output averaged patterns for excitations in the center of the Brillouin zone and at the Dirac point disappears also upon narrowing of the input envelope at fixed disorder level [Fig. 5(b)]. With decrease of $w$ the spatial spectrum of the input excitation broadens, since its characteristic width is proportional to $2\pi/w$. For sufficiently small $w$ values the width of the spectrum becomes comparable with the width of the Brillouin zone, and the unusual diffraction features characteristic of such isolated points as Dirac ones disappear. Still, for all widths $w$ the localization is more pronounced for the excitation centered at $k_\eta, k_\zeta = 0$. For large values of $w > 10$ the form-factor changes as $\chi_{av} \sim 1/w$.

In summary, by comparing the evolution of excitations in the center of the Brillouin zone and at the Dirac point in disordered honeycomb lattices we observed substantial differences in the widths of localized patterns at small disorders. Such differences are connected with different expansion rates for normal and conical diffraction.

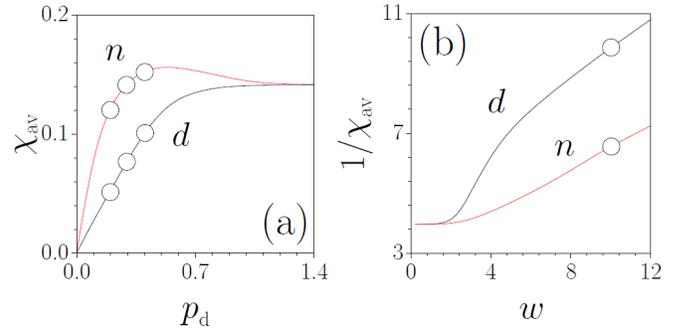

Fig. 5. (a) Averaged integral form-factor versus disorder level $p_d$ at $w=10$. (b) Inverse averaged form-factor versus width of the input envelope $w$ at $p_d=0.4$ (b). The lines marked with letter "$n$" correspond to the excitation at the top of the first allowed band, while lines marked with "$d$" correspond to the excitation in the Dirac point. Circles correspond to the averaged intensity distributions shown in Fig. 3.